# Docker under Siege: Securing Containers in the Modern Era

Gogulakrishnan Thiyagarajan[1], Prabhudarshi Nayak[2]

[1]*Software Engineering Technical Leader, Cisco Systems Inc, Austin, Texas, USA.*
[2]*Faculty of Engineering and Technology, Sri Sri University, Cuttack, Odisha, India.*

Corresponding Author: gogs.ethics@gmail.com

## Abstract

*Containerization, driven by Docker, has transformed application development and deployment by enhancing efficiency and scalability. However, the rapid adoption of container technologies introduces significant security challenges that require careful management. This paper investigates key areas of container security, including runtime protection, network safeguards, configuration best practices, supply chain security, and comprehensive monitoring and logging solutions. We identify common vulnerabilities within these domains and provide actionable recommendations to address and mitigate these risks. By integrating security throughout the Software Development Lifecycle (SDLC), organizations can reinforce their security posture, creating a resilient and reliable containerized application infrastructure that withstands evolving threats.*

## Introduction

Containerization has brought a sea change in application development and deployment paradigms. Docker is one of the most prevalent platforms, ensuring that the management of containerized applications becomes lightweight and practical. However, the rapid proliferation of Docker has raised several security challenges that need to be tackled by an organization to keep applications safe from vulnerabilities and cyberattacks. Recent discussions on container security have underscored several challenges, including the necessity for enhanced visibility into container activities, insufficient expertise among teams, and inadequate collaboration between security and development groups. (IANS, 2022). Besides this, the complex natures of container environments often make traditional security ineffective. This paper reviews critical aspects of container security, which include image security, runtime security, network security, configuration management, supply chain security, and monitoring and logging. It underlines different shortcomings in the state-of-art Docker security practices and provides concrete mitigation strategies. Incorporating end-to-end security within the SDLC significantly improves organizations' overall security posture in a containerized environment. Literature Review & Related Work As a result of rapid growth in the adoption of containerized environments, mainly because of Docker, container security has turned hot. Several researchers focused on different areas of securing containers, which include, but are not limited to, image security and runtime monitoring. Some critical contributions regarding the same are briefly summarized below:

| Study | Focus Area | Methodology | Key Findings |
|---|---|---|---|
| Merkel (2014) | Container isolation and efficiency | Introduced Docker containers for application isolation and deployment. | Highlighted the advantages of containerization but noted potential risks in isolation mechanisms. |
| Shin & Park (2018) | Container Security Tools | Introduced Docker containers for application isolation and deployment. | Identified gaps in existing security tools, particularly in runtime protection. |
| Casalicchio (2019) | Orchestration & Security in Docker | Analyzed orchestration tools like Kubernetes with a focus on security. | Suggested enhancements in Kubernetes' security features, especially in resource allocation. |
| Sultan et al. (2019) | Container Supply Chain Security | Investigated supply chain vulnerabilities, focusing on third-party libraries | Found that 70% of security issues in containers are linked to unvetted dependencies from public sources. |
| Zhang & Liu (2020) | Runtime Security for Containers | Reviewed runtime security challenges and proposed anomaly detection techniques. | Found that runtime misconfigurations are a major vulnerability that can be exploited for attacks. |
| McLaughlin (2021) | Docker Security Best Practices | Practical guide on Docker security, focusing on image verification and least privilege principle. | Suggested the use of automated tools for continuous image scanning to mitigate supply chain risks. |
| Alyas et al. (2022) | Vulnerability Management | Demonstrates practical tools and methods for vulnerability management, aligning with image security | Proposed a system for managing container vulnerabilities using Docker Engine, addressing performance and security challenges. |

*Summary*

**Docker Under Siege: Securing Containers in the Modern Era**, explores the rapid rise of containerization in application development and deployment, particularly focusing on Docker. Containerization offers clear advantages, such as efficiency, scalability, and portability. However, the widespread adoption of Docker has brought forth numerous security challenges that organizations must address to ensure the reliability and resilience of containerized applications.

This study delves into critical areas of container security, including the protection of container runtime, network security, configuration management, and supply chain vulnerabilities. The paper also emphasizes the need for continuous monitoring and logging as a means to detect and address security threats in real time. Each of these areas presents distinct security challenges: for instance, base image vulnerabilities, misconfigurations in runtime, insufficient network segmentation, and the risks of using third-party libraries and dependencies from public repositories.

To counter these issues, the document provides actionable recommendations, advocating for best practices in security that extend throughout the software development lifecycle **(SDLC)**. This involves implementing strict access controls, regular vulnerability scanning, and adherence to the least privilege principle to limit the impact of potential breaches. The study also underscores the importance of maintaining updated and verified base images, strengthening runtime protections, and enforcing network segmentation policies to secure sensitive data.

Through a combination of practical strategies and robust governance measures, organizations can fortify their security posture in containerized environments. This paper serves as a guide for enhancing Docker container security, laying out a framework that can help organizations protect their applications against evolving cyber threats. By integrating these security measures into development workflows, organizations can achieve a resilient, scalable, and secure container infrastructure that supports modern application needs while safeguarding against vulnerabilities inherent in containerized environments.

## 1. Image Security: The Importance of Securing Docker Images

The security of Docker containers fundamentally depends on their base images. These base images serve as a blueprint for the Docker environment's contents, including essential libraries, dependencies, and software the application needs to run. If the integrity of the container originates from the base image from which it was derived, then any breach of the base image could translate into a massive vulnerability in its resulting container. Insecure or outdated base image use can leverage attacks by vulnerabilities in the containerized environment. A vulnerable image may carry an insecure configuration, malware, or any other vulnerability that may compromise the application running in the container and the underlying host system. Therefore, base images must be secure and updated to maintain a sound security posture.

Containerization in modern software development has brought many benefits, such as greater scalability, portability, and resource efficiency. However, this technology has also introduced several significant security challenges. One of the most pressing concerns related to the usage of containers today is the reliance on images from untrusted sources. This is very common in the case of organizations that pull images from publicly available repositories without proper due diligence- a practice that poses several security risks. These risks can compromise the integrity of the container ecosystem and undermine its general security posture.

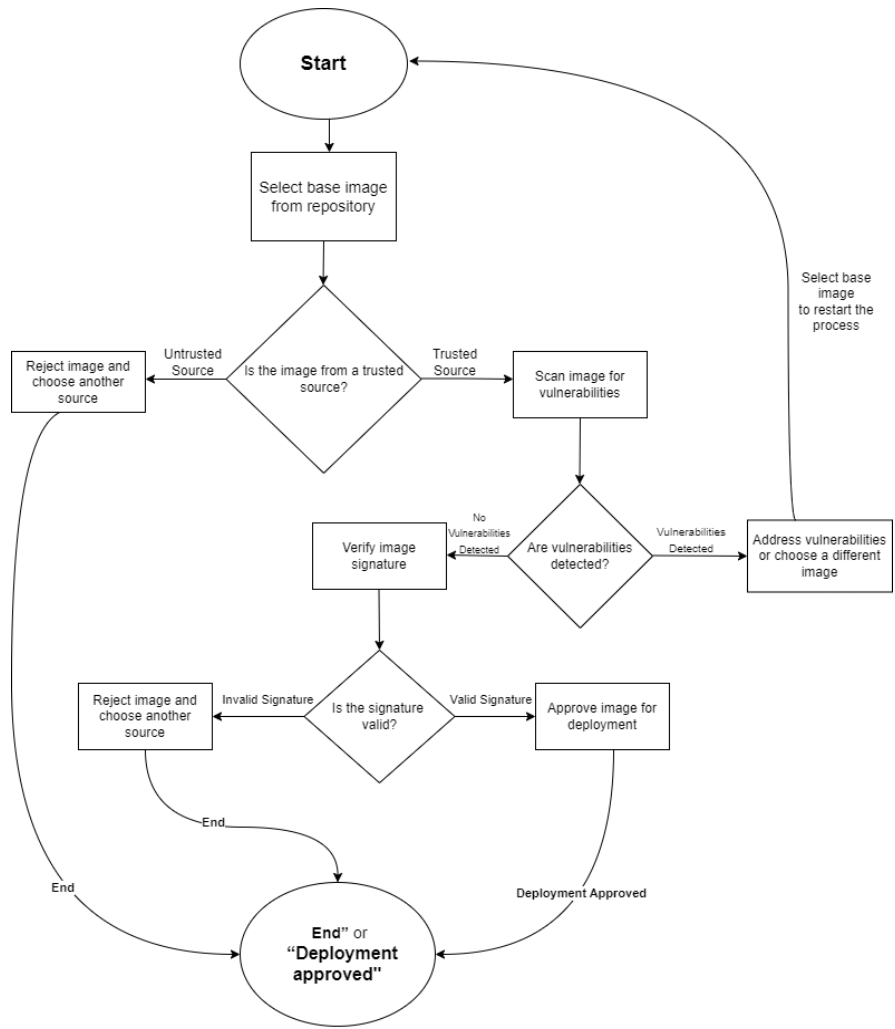

Image Scanning and Verification Process flowchart

**Fig1: Image Scanning and Verification Process flowchart**

The **Image Scanning and Verification Process** flowchart is a critical visual aid for understanding how to ensure container security before deployment. The process begins with selecting a base image from a repository and proceeds through a series of checks and scans designed to verify the image's integrity and security. Below, each step of the flowchart is described in detail:

1.**Start:** The process initiates at the starting point, represented by an oval labeled "Start." This symbol indicates the commencement of the image selection and verification workflow.

2.**Select Base Image from Repository:** The first step involves choosing a base image. This is represented by a rectangle labeled "Select base image from repository." Here, developers or security teams pick an image that serves as the foundation for containerized applications.

3.**Is the Image from a Trusted Source?:** The selected image must be evaluated for its source's reliability. This decision point is depicted by a diamond labeled "Is the image from a trusted source?" It branches into two possible paths:

- If No: If the image is not from a trusted source, it should be immediately rejected. The pathway leads to a rectangle labeled "Reject image and choose another source," followed by an "End" oval to indicate the conclusion of the process at this stage.

- If Yes: If the image is verified as being from a trusted source, the process continues to the next step.

4. **Scan Image for Vulnerabilities**: The next step involves scanning the image for known vulnerabilities. This is shown as a rectangle labeled "Scan image for vulnerabilities." This step is crucial for identifying any existing flaws that could compromise the container's security.

5. **Are Vulnerabilities Detected?:** After scanning, the process reaches another decision point, represented by a diamond labeled "Are vulnerabilities detected?" This step determines whether any security issues were found.

- If Yes: If vulnerabilities are detected, the pathway leads to a rectangle labeled "Address vulnerabilities or choose a different image." This may involve fixing the issues or starting over by selecting a new base image, which loops back to the initial "Select base image" step.

- If No: If no vulnerabilities are found, the process continues to the next verification step.

6. **Verify Image Signature**: This step ensures that the image has not been tampered with and that it originates from a legitimate source. It is represented by a rectangle labeled "Verify image signature." Verifying the image signature is crucial for confirming its authenticity and integrity.

7. **Is the Signature Valid?**: The signature verification results are assessed at this decision point, shown as a diamond labeled "Is the signature valid?"

- If No: If the signature is invalid, the image should be rejected, as indicated by a rectangle labeled "Reject image and choose another source." This leads to an "End" oval, ending the process.

- If Yes: If the signature is valid, the process moves forward.

8. **Approve Image for Deployment**: If the image passes all checks and scans, it is approved for deployment. This final action is represented by a rectangle labeled "Approve image for deployment."

9. **End:** The process concludes with an "End" oval labeled "Deployment approved," signaling that the image is now ready for use in the containerized environment.

## 1.1 Current Gaps

- **Untrusted Sources**:

Using Docker images from untrusted sources is highly dangerous because of malicious code and many other vulnerabilities. Most Docker images are based on various open-source libraries and packages, which may contain known security vulnerabilities. This has been reflected in multiple studies that revealed that even the widely acknowledged Docker images can hide many vulnerabilities; some reports show that popular images could have more than 30 known security flaws [1][2]. That again underscores the importance of thorough reviews before any Docker image is deployed into production.

As a best practice, an organization should take several measures to mitigate the associated risks from untrusted Docker images. First, only official and verified images from trusted sources, such as Docker Hub, should be utilized to minimize exposure to known vulnerabilities [1][2]. By implementing an aggressive scanning process through Snyk, one can identify and address security issues in the images before deployment [2]. Using multi-stage builds is another good practice that ensures development dependencies do not make their way to production images, inadvertently increasing the attack surface[2]. Another suggestion for organizations is to avoid generic tags like "latest" instead of image version tags. That offers a way to ensure reliability and consistency so that when changes occur, an unexpected shift to the base image does not bring in new vulnerabilities [1][2]

Finally, organizations should regularly update the Docker images and enforce image access management to ensure which images are deployed within their environments. Considering this suggestion will enable companies to vastly improve their security posture against threats emerging from untrusted Docker images[1][2]

- **Unknown Vulnerabilities**:

The sudden embracing of container technology in general, and Docker in particular, changed the deployment and management of applications. Conversely, it brought huge security risks for organizations, particularly those dealing with unknown vulnerabilities in Docker images. These could include outdated libraries, unpatched software, or misconfigurations not found during setup time. A study conducted by Malhotra et al. underscores the significance of assessing vulnerabilities in Docker Hub images, indicating that numerous official and verified images contain security flaws attributed to inadequate monitoring and infrequent updates[3].

The complexity of unidentified vulnerabilities in the Docker ecosystem is such that images consist of several layers, each of which may pose a different risk. Without regular scanning and assessments for vulnerabilities, an organization may deploy an image that can open its systems to vulnerabilities. Tools like Trivy and Clair are crucial in finding these vulnerabilities, but they often rely on a database of known issues, which means newly discovered or less common ones fall through the cracks[4]. For this reason, proactive management of vulnerabilities is required: images should be scanned with certain periodicity and updated according to the newly emerging threats.

Dependence on third-party libraries in Docker images could also increase the chances of vulnerability exploitation. While pulling images from public repositories, organizations may incidentally introduce specific vulnerabilities related to those libraries into their environments, mainly if they are outdated or poorly maintained. Protection against such possible exploitation requires continuous security posture monitoring and assessment of the Docker image[4]. Organizations can significantly reduce the risks associated with unknown vulnerabilities by implementing a strategy incorporating vulnerability scanning, dependency management, and strict policies on image usage.

In conclusion, the more Docker gains significant attention in application deployments, the more an organization should be able to understand and manage unknown vulnerabilities. Strong vulnerability scanning practices, frequent image updates, and a culture of security awareness are necessary to mitigate the risks of using Docker images in production environments[4].

A significant concern is the dependence on obsolete dependencies. Container images may incorporate un-updated libraries or software packages, notwithstanding well-documented vulnerabilities. For instance, vulnerabilities in widely utilized libraries such as OpenSSL (CVE-2020-1971) or PHP-FPM (CVE-2019-11043) are frequently disregarded, creating critical security gaps. Furthermore, improperly configured permissions within containers represent a prevalent vulnerability. Running containers as root or setting too permissive file permissions-such as chmod 777- provide opportunities for attackers to escalate privileges and run malicious activities. Known vulnerabilities like the Linux Kernel vulnerability CVE-2021-22555 and the Dirty Pipe exploit CVE-2022-27666 point to misconfigurations.

Besides that, improper network configurations in containers can expose sensitive services to unauthorized users or allow denial-of-service attacks. Compassionate cases involve exposed API endpoints and open container ports, as seen in CVE-2018-1002105 in Kubernetes and CVE-2020-10749 in Docker Daemon. Another critical issue is hard-coded secrets. This means the developers explicitly store API keys, passwords, or encryption keys within the container image. This class of vulnerabilities is hazardous since the attacker immediately obtains access to critical systems. A good example is the MongoDB credentials leak (CVE-2021-31684), which underlines one of the consequences of poorly handled secret management.

Base images, if unpatched, are a significant cause of the increased severity of the issue. Often, containers are built on base images of older versions of Ubuntu or Alpine Linux that don't get frequent updates regarding security vulnerability patches. For instance, consider CVE-2020-11444 (Ubuntu Image) and CVE-2019-5021 (Alpine Linux), which are examples of vulnerabilities that persist due to a lack of updates in base images. Beyond those issues, containers escape vulnerabilities-the most critical risks are CVE-2019-5736 in runc and CVE-2016-9962 in Docker. These will allow an attacker to break out of the container's isolated environment and give them access to the host, thus threatening the entire infrastructure.

Image poisoning and dependency confusion are both growing risks in the world of cybersecurity. Image poisoning occurs when hackers upload compromised or altered container images to public registries, which are then unknowingly deployed by developers. For example, CVE-2018-20685 Docker Image Signature Bypass shows how tainted images can bypass security checks. On the other hand, with dependency confusion, the attackers inject malicious packages into public repositories that become part of the container builds. A severe reminder of the damages it could impose is given by the well-known Log4j vulnerability CVE-2021-44228.

| Vulnerability Type | Description | Common Vulnerable Code | Relevant CVEs | Severity |
|---|---|---|---|---|
| Outdated Dependencies | Many container images rely on outdated libraries or software packages that have known vulnerabilities but haven't been patched. | - Using old versions of libraries like OpenSSL, Java, or Python packages. | CVE-2020-1971 (OpenSSL), CVE-2019-11043 (PHP-FPM) | High |

| | | - Neglecting to update base images. | | |
|---|---|---|---|---|
| Misconfigured Permissions | Incorrect permission settings within container images can allow unauthorized access to files, directories, or system functions. | - Setting overly permissive file permissions (e.g., chmod 777).<br>- Running containers as root instead of a restricted user. | CVE-2021-22555 (Linux Kernel), CVE-2022-27666 (Dirty Pipe) | Critical |
| Insecure Network Configurations | Misconfigured network settings in containers may expose sensitive services to the internet, leading to unauthorized access or denial-of-service attacks. | - Exposing container ports to the public without proper firewall rules.<br>- Unsecured API endpoints. | CVE-2018-1002105 (Kubernetes), CVE-2020-10749 (Docker Daemon) | High |
| Hardcoded Secrets | Sensitive information, such as API keys, database credentials, or encryption keys, is hardcoded directly into container images. | - Hardcoding database passwords in Dockerfile or environment variables.<br>- Insecure storage of API keys in application code. | CVE-2021-31684 (MongoDB Credentials Leak), CVE-2020-0601 | Critical |
| Unpatched Base Images | Vulnerabilities exist in the base images on which containerized applications are built, which often need to be updated regularly. | - Using base images like ubuntu:14.04 without security updates.<br>- Using unmaintained or vulnerable open-source images. | CVE-2020-11444 (Ubuntu Image), CVE-2019-5021 (Alpine Linux) | High |
| Container Escape Vulnerabilities | Attackers exploit flaws that allow them to break out of the container's | - Vulnerabilities in container runtime | CVE-2019-5736 (runc), CVE- | Critical |

| | isolated environment and access the host system. | engines (e.g., Docker, runc).<br>- Weak namespace isolation. | 2016-9962 (Docker) | |
|---|---|---|---|---|
| Image Poisoning | Malicious actors may compromise or tamper with publicly available container images, injecting malicious code or backdoors. | - Uploading trojanized images to public registries.<br>- Using unverified images from unofficial sources. | CVE-2018-20685 (Docker Image Signature Bypass), CVE-2020-8910 | High |
| Dependency Confusion | Attackers inject malicious packages into publicly accessible libraries, which are unintentionally pulled into container builds. | - Accidentally pulling dependencies from untrusted repositories.<br>- Using npm install or pip install without verification. | CVE-2020-26233 (npm), CVE-2021-44228 (Log4j) | Critical |

- **Malicious Code**

Malicious code embedded in Docker images severely threatens container security. Therefore, the vulnerabilities of official and verified images on Docker Hub must be assessed. Recent research by Malhotra, Bansal, and Kessentini has underlined the urgency of performing a comprehensive vulnerability analysis on these images since they are often used as a starting point for deploying applications into production environments[5]. The researchers' study employed several open-source vulnerability detection tools to analyze the security posture of those images and found that many images contain critical vulnerabilities that attackers could potentially exploit. The finding has shown that rampant outdated libraries and poor security practices in commonly used images raise the possibility of malicious code execution, hence the requirement for sound security practices in managing containers[5].

| Type of Malicious Code | Description | Example | Mitigation |
|---|---|---|---|
| ***Backdoors*** | Hidden access methods to bypass security controls. | SSH backdoors or hard-coded credentials. | Use trusted sources and signature verification. |

| *Ransomware* | Encrypts files, demanding a ransom for decryption. | Encrypts critical data after deployment. | Regular backups, offline scanning |
|---|---|---|---|
| *Trojan Horses* | It appears legitimate but contains hidden malicious functions. | Web server container with reverse shell code. | Static analysis and secure coding practices. |
| *Cryptojacking Scripts* | Steals system resources for unauthorized cryptocurrency mining | Mining cryptocurrency in the background affects performance. | Resource monitoring, scanning images. |
| *Rootkits* | Provides unauthorized root access while hiding its presence. | They compromised the Linux kernel in the container. | The least privilege principle is runtime monitoring. |
| *Supply Chain Attacks* | Malware is introduced via compromised dependencies or images. | Malicious third-party library in the container stack. | Regular updates and dependency validation. |
| *Botnets* | Containers are used to form part of a network for large-scale attacks. | Container connecting to a command-and-control server | Network segmentation, runtime behavior monitoring. |
| *Information Stealers* | Targets sensitive data like credentials or environment variables. | Extracts AWS keys from running containers. | Encrypt sensitive data, secrets management. |

- **Lack of Accountability**

Lack of accountability within container ecosystems is one of the most critical security challenges, especially regarding Docker images. As containerization gains more and more popularity, the integrity and security of container images are usually maintained by users rather than providers. This fact is amplified by the reality that many developers rely on images derived from public repositories, such as Docker Hub, without having a full-fledged idea of their lineage and the security measures adopted during their development. Malhotra et al. (2023) point out that easy access to untrusted and unverified images increases the risk of vulnerability introduction, as such photos could contain nefarious code or rely on outdated libraries[6]. The intrinsic anonymity in creating and distributing container images leads to a lack of accountability, making it difficult to track where vulnerabilities come from whenever there is a security breach.

Moreover, the decentralization of container technology exacerbates issues of accountability. Unlike traditional software deployment, which updates and patches from one central authority, images could be easily pulled out and deployed from several sources. Fragmentation like this will mean security vulnerabilities persist for years, as there is no central authority to monitor and maintain the image set. Jiang and Zheng (2020) articulate that the intrinsic design of container systems, which facilitates swift iteration and deployment, frequently overlooks security considerations, necessitating that users independently manage associated risks[7]. Consequently, this absence of accountability not only endangers the security of individual applications but also diminishes the overall trustworthiness of containerization as a practical deployment strategy.

Improved governance and accountability frameworks in the container ecosystem should be developed to address these challenges. This suggests tightening the verification processes of images uploaded to public repositories and leveraging automated scanning tools for photo vulnerabilities. Finally, official, verified pictures should be used, and policies should be established to audit and update container deployments on a regular schedule. Some of the risks of containerization can be mitigated with accountability within the developer and organizational culture, leading to a more secure operational environment.

*1.2 Recommendations*

- **Securing Image Origins**:

Organizations should ensure that official and verified images are used and sourced from trusted registries like Docker Hub to reduce the risks related to untrusted sources actively. This can be ensured by implementing DCT, which ensures that only signed images can be deployed, increasing the deployed containers' integrity [8]. Moreover, it should be an organization's policy to ensure strict denial of image usage from unknown or unverified sources, reducing the likelihood of a security breach to a great extent. The regular training and awareness programs for the development and operations teams help inculcate a security-oriented culture where the reasons for choosing trusted images and associated risks with them are taught[9]. This proactive step helps reduce risks and promotes best practices for container security.

- **Addressing Unknown Vulnerabilities**:

Addressing unknown vulnerabilities requires a proactive vulnerability management strategy. An organization needs to regularly scan Docker images for vulnerabilities with scanners, such as Trivy, Aqua Security, or Clair, capable of detecting and remediating potential security risks before they can be exploited[10][11]. Keeping an updated inventory count of the Docker images and monitoring them regularly will ensure that organizations are aware of new emerging threats. Moreover, the periodic execution of security audits and penetration tests will reveal hidden vulnerabilities so that teams can take proactive action to eliminate defects. Establishing an update policy for all Docker images, including applying patches in time, is crucial for protecting the container environment against vulnerabilities [12].

- **Mitigating Malicious Code Threats**:

To mitigate malicious code attacks, an organization should have a multi-layered approach towards security. It should conduct rigorous scanning of all the images before deployment. Also, it leverages static and dynamic analysis tools to identify malicious code hidden inside the pictures [13]. Additionally, best practices, such as running containers with the principle of least privilege and restricting sensitive resource access, can significantly reduce the impact of any malicious code that might be present. Continuous education and awareness about malicious code risks will help teams proactively identify and mitigate the threats [14].

- **Accountability Improvement in Container Management**:

Accountability within a Docker ecosystem is paramount in enforcing sound security governance. Implementing RBAC ensures only the right people can deploy or change containers[15]. To allow accountability, comprehensive logging of all activities related to pulling images, deployment, and changes is essential in tracking the actions of individual persons or teams. Regular audits and compliance checks reinforce security policies and best practices, increasing accountability within an organization. Team members need to be encouraged to develop a sense of responsibility to ensure that security standards are maintained and the risks from container usage are minimized [16].

## 2. Runtime Security: Importance of Runtime Security

Runtime security forms the most critical aspect of a containerized environment. Containers are highly dynamic, where an application or service runs in a secured environment, though only partially flawless. The runtime is essential to protect the containers against several risks, such as control compromise, data leakage, or service disruption. They could be exposed to runtime misconfigurations, breached application code vulnerabilities, or weaknesses in the underlying infrastructure on which the container is operating. Unless adequately secured, attackers could compromise the containerized environment, and leakage of sensitive data would result in service disruptions, financial loss, and damage to one's reputation. Ensuring integrity, confidentiality, and availability of containers at runtime is of utmost importance. **(Flauzac2020).**

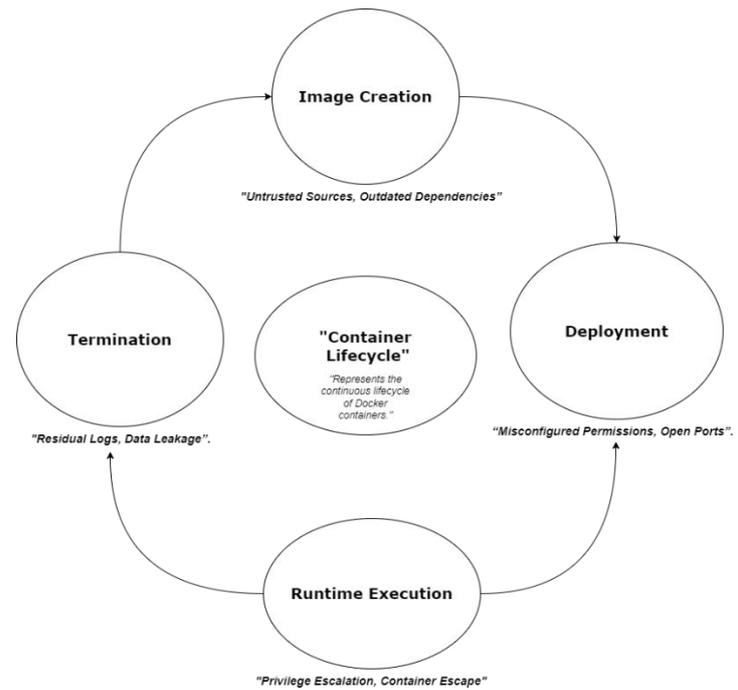

*Fig2: Container Lifecycle and Runtime Security Risks*

### 2.1 Current Gaps

- **Over-Privileges:**

Over-privileges within Docker containers create a huge security risk that might eventually compromise the integrity of the containerized applications and host systems. This includes privilege leakages around containers being granted excessive permissions than needed for operation, enabling them to perform sensitive operations or access restricted resources on the host machine[17]. Over-privileged configurations may present potential vulnerabilities since attackers can leverage these permissions to carry out unauthorized actions, elevate privileges, or even compromise the host system. Besides, such security challenges are further exacerbated because of the shared nature of the Linux kernel across containers, which can facilitate lateral movement across the container ecosystem if the attacker manages to gain access to one over-privileged container. Consequently, organizations must manage the permissions assigned to Docker containers meticulously, employing the principle of least privilege to mitigate risks and enhance overall security[17]. Addressing over-privileges is essential for establishing a secure and resilient container infrastructure, safeguarding sensitive data and preserving operational integrity within cloud environments.

- **Insufficient Isolation:**

Insufficient isolation of Docker containers is a critical security risk because the Docker containers share an operating system kernel with the host. A vulnerability or attack against one container due to weak separation may easily compromise the host and affect other containers negatively, leading to severe risks such as container escape[18]. Attackers will leverage one single container misconfiguration or vulnerability to access sensitive data on other containers or the host environment. This increases the attack surface. Moreover, poor isolation will challenge security policies or regulatory requirements, so organizations seek to adequately secure sensitive information[18]. The dynamic and transient nature of containers makes managing security consistently even more challenging. To address these issues, organizations must implement rigorous security hardening practices. Such practices include using namespaces and control groups to enforce improved isolation, thereby reducing the likelihood of successful attacks and enhancing the overall security of containerized environments.

- **Insecure Runtime Behavior:**

Insecure runtime behavior by Docker containers poses serious security challenges, driven primarily by the intrinsic nature of containerization and the complicated interactions with the host operating system. At runtime, a container is exposed to various vulnerabilities, from insecure configurations and lack of access controls to the download and execution of malicious code[19]. This is all the more an issue in that the dynamic nature of containers further exacerbates these issues since vulnerabilities can quickly propagate in containerized environments if not acted upon immediately[20]. Moreover, the containers usually run with high privileges or over permissive roles, which exposes them to unauthorized access or, worse, privilege escalation that may cause a significant security incident[20]. Security controls at the minor privilege level, effective runtime monitoring, and periodic vulnerability assessment should be implemented strictly to ensure that any deviation from expected behavior is reliably detected and contained[19]. In a nutshell, insecure runtime behavior is one of the most critical issues that must be addressed while maintaining strong security in Docker environments against possible exploits seeking to compromise sensitive data and application integrity.

- **Poor runtime integrity monitoring:**

Poor runtime integrity monitoring in Docker containers carries significant risks to the security and reliability of applications. Without proper monitoring mechanisms, malicious actions could be performed without detecting possible breaches or exploiting vulnerabilities within containers[21]. Runtime integrity monitoring is essential for the detection of unauthorized changes in application code and the environment, making sure that only legitimate processes are running inside the container. Without such safeguards, compromised containers can be manipulated to perform malicious activities and jeopardize the overall integrity of the host system and other containers running on it. Moreover, the current approaches to monitoring lack comprehensive solutions that provide real-time insight into the integrity of containerized applications, exposing organizations to attacks exploiting these monitoring gaps[21]. As containerization becomes increasingly prevalent in cloud environments, improving runtime integrity monitoring is essential to uphold security standards and protect critical data and systems from emerging threats.

## 2.2 Recommendations

- **Least Privilege Principle:**

The Least Privilege Principle in Docker containers is a fundamental security practice that emphasizes granting containerized applications only the minimum permissions necessary to perform their designated functions. This principle significantly reduces the attack surface by limiting the potential for unauthorized access or malicious activities within the container environment[22]. Organizations can mitigate the risks associated with privilege escalation attacks by ensuring that containers operate with restricted privileges, where attackers seek to gain elevated access to resources or execute harmful commands. The advantages of implementing the Least Privilege Principle include enhanced security posture, improved compliance with regulatory requirements, and an overall reduction in the scope of security incidents. Furthermore, this approach streamlines incident response efforts since compromised containers are less likely to access sensitive data or critical infrastructure, thereby effectively containing potential breaches. Adopted practices often involve using container orchestration tools to enforce strict access controls and automating privilege auditing, ensuring that containers remain compliant with the principle of least privilege throughout their lifecycle[23]. Adhering to this principle is vital for robust security in Docker environments, enabling organizations to operate with increased confidence in their containerized applications.

- **Improved Isolation:**

One of the major improvements is better isolation in Docker containers, which further improves the security and efficiency of containerized applications. This enhanced isolation creates a secure boundary between the host operating system and each container, drastically reducing the risk of unauthorized access and potential cross-container vulnerabilities[24]. Some benefits of better isolation include improved security posture, as it reduces the potential for attacks propagating from one container to another or affecting the host system. Better isolation supports compliance with regulatory standards, enabling an organization to protect sensitive data more effectively. Further, it allows for more predictable application behavior since each container runs in its environment with dependencies, reducing conflict and instability. This enhanced architecture supports scalability and resource management, allowing organizations to run multiple containers efficiently without compromising performance. Improving isolation is critical to maintaining secure, reliable, and efficient operations in Docker environments. This will pave the way for increased trust and flexibility in deploying cloud-native applications.

- **Runtime Monitoring and Intrusion Detection:**

The most central aspects of security and operational integrity in containerized environments are runtime monitoring and intrusion detection within Docker containers. This includes continuous observation of activities, system calls, resource utilization within the containers, and any deviation that may signal a security breach or unexpected failure in real time[25]. Effective runtime monitoring and IDS have brought down the benefits of enhancing threat detection capabilities, enabling organizations to respond quickly to potential intrusions to mitigate risks in real time[25][26]. Active system behavior monitoring informs organizations about performance issues, resource bottlenecks, and application vulnerabilities, enhancing overall system reliability[25]. Moreover, with the dynamic nature of containerized applications, more than traditional security may be required. Therefore, having an integrated runtime monitoring strategy allows organizations to adapt to emerging threats and comply with the best security practices [25][26]. Lastly, effective runtime monitoring and IDS contribute to a robust security posture, ensuring containerized applications can operate safely and efficiently in complex, cloud-based environments.

- **Runtime Security Tools Usage:**

Runtime security tools within Docker containers are significant in protecting containerized applications and ensuring operational integrity throughout the application lifecycle. The tools provide real-time monitoring and alerting mechanisms; such features help detect security threats and anomalies during runtime and enhance the security posture of the container environment[27]. An essential advantage of using runtime security tools is detecting and preventing possible vulnerabilities before they become exploitable, significantly reducing the chance of data breaches and service disruptions[28]. Container operations are secured with tools like Falco and Cilium Tetragon, allowing for dynamic policy enforcement, system call monitoring, and tracking of container behavior to respond to any suspicious activity[28] quickly. Integrating advanced solutions like eBPF increases performance and enhances contextual awareness for security monitoring, resulting in lower false-positive rates and improved detection accuracy[29]. With this sort of robust security measure, organizations are safeguarding sensitive data and improving their compliance levels regarding various industry regulations, enhancing trust with stakeholders and clients. The strategic application of Runtime Security tools enhances the resiliency of the framework, making it possible to deploy and manage Docker containers safely in increasingly complex environments.

- **Integrity Checks and Image Validation:**

Integrity checks and image validation are essential for securing and ensuring the reliability and consistency of containerized applications in Docker containers. This, therefore, means that once created, Docker images should not be modified or tampered with; this guarantees that the software runs in a secure state[30]. Integrity checks and image validation bring in the benefits of better security, as organizations will be able to proactively identify any unauthorized changes or potential vulnerabilities before deploying images into production[30]. Further, teams ensure that the container images come from trusted sources and meet all compliance standards—something significant for regulated industries[30]. With such checks, the usage of corrupted or outdated images is prevented, contributing to the general stability and performance of applications and, therefore, results in more reliable deployments with less downtime[31]. Integrity checks and image validation may be part of workflows that provide an organization with a more robust security posture and make its deployment processes streamlined and efficient for containerized applications to be secure and efficient.

- **Runtime Network Security:**

Runtime network security in Docker containers is essential for protecting containerized applications from a long line of cyber threats that operate when the application is up and running. This security framework involves monitoring and managing the network traffic between containers and between containers and external systems to prevent unauthorized access and attacks such as Man-in-the-Middle (MitM) or network-based exploits[32]. One of the most compelling benefits of running strong runtime network security is the protection it can provide against vulnerabilities resulting from shared networking environments, where numerous containers work off the same host[32]. Firms will have to execute mechanisms like FWC or isolation methods for creating secure gateways that ensure effective risk mitigation and traffic filtering, enhancing the overall security posturing of a container environment[33]. Such security measures also help adhere to regulatory requirements since they offer a systematic approach to monitoring and protecting sensitive data. Runtime network security allows for more excellent operational stability, ensuring that the services remain uninterrupted and resilient against potential attacks—thus giving clients and stakeholders increased assurance in the security of the deployed applications[34]. In a nutshell, runtime network security is an essential strategy that should be implemented to ensure the integrity and safety of Docker containers within the current digital ecosystem.

## 3. Network Security: Importance of Network Security

In contemporary containerized environments, the network facilitates secure communication among services. Containers are frequently deployed across distributed environments, encompassing multiple hosts or cloud infrastructures. This escalation in deployment complexity heightens the challenges associated with securing the network and generates numerous points of vulnerability if not adequately managed. A network security breach may further lead to unauthorized access, exfiltration of data, or even the compromise of an entire application stack. According to Merkel (2014), ensuring the security of communications between containers is crucial in maintaining an organization's overall security posture. **(IANS, 2022).**

### 3.1 Current Gaps

- **Permissive Communication:**

Permissive communication in container security is a significant concern that addresses the vulnerabilities resulting from inadequate network policy configurations in containerized environments. When containers can communicate freely without stringent network restrictions, it can lead to various security risks, including unauthorized data access and lateral movement by attackers within the network. This unrestricted access allows malicious actors to compromise a single container and navigate to other containers, potentially leading to widespread breaches across applications and data repositories. The ease of such lateral movement not only exacerbates the impact of a security breach but also complicates the enforcement of the principle of least privilege, which suggests that systems should limit access to essential functions only. [35] Organizations must implement stricter network segmentation to combat these challenges and define precise communication policies that allow only essential interactions between trusted containers. Such measures are crucial for safeguarding sensitive information and maintaining a robust security posture as organizations increasingly adopt containerization technologies. [36] Ultimately, addressing permissive communication will play a vital role in enhancing the overall security of containerized applications and mitigating the risks posed by evolving cyber threats.[37]

- **Lack of Encryption:**

It cannot be gainsaid that open network communications pose severe risks in light of evolving cyber threats. Without proper encryption mechanisms, sensitive data transmitted over the internet are liable to be intercepted by miscreants of various kinds, leading to unauthorized access to systems and data breaches. This vulnerability is highly concerning for IoT devices, which often operate on minimal security levels and can easily be manipulated to compromise entire networks. Effective encryption strategies have become paramount in safeguarding data integrity and confidentiality. They ensure that they remain incomprehensible to unauthorized users even when data could get intercepted. [38]

| Issue | Explanation | Type of Encryption Suggested |
|---|---|---|
| Lack of Encryption in Containers | In many containerized environments, traffic between containers is often unencrypted, exposing systems to risks such as MITM (Man-in-the-Middle) attacks. | TLS (Transport Layer Security) |
| Unprotected Data During Transit | Sensitive data such as authentication credentials, API keys, and other private information can be intercepted during transmission. | SSL/TLS Encryption for HTTPS |
| Multi-Cloud & Hybrid Environments | This issue is even more critical in multi-cloud or hybrid setups where data moves across various networks, some of which may not be fully secure. | IPSec (Internet Protocol Security) or VPN (Virtual Private Network) encryption |

**Key Points:**

- Unencrypted network traffic in containers can lead to the exposure of critical data.
- Utilizing encryption mechanisms like TLS or SSL can safeguard sensitive information.
- Multi-cloud or hybrid environments, where data travels through multiple networks, are particularly vulnerable and require additional encryption layers such as IPSec.
- **Dynamic and Ephemeral Nature of Containers:**

The containers' most characteristic dynamic and transient nature significantly raises their utility in modern cloud computing. Thus, container creation, scaling, and destruction can be performed incredibly quickly, enabling organizations to deploy applications with increased agility and flexibility. [39] This ability to dynamically adjust resources allows businesses to respond rapidly to changing demands while optimizing resource utilization. [40] The temporary nature of the containers will enable them to live for a short period or just until they achieve what they were meant to and ensure an application is light and replaceable without affecting system stability. This permits smooth integration and continuous deployment, CI/CD, enabling development teams to introduce innovations with the least shutdowns possible. Container lifecycles become essential to understand as more organizations adopt container orchestration tools to manage these dynamic workloads, which will be crucial for having robust cloud strategies.

- **Limited Network Monitoring and Logging:**

The limited network monitoring and logging in containerized environments create immense challenges in maintaining robust security and operational efficiency. This is because the fast-moving flows within the network might challenge the traditional monitoring tools and the transient lifestyles of containers. [41] This lack of visibility can further lead to undetected anomalies that make identification and response to emerging security threats or performance issues quite complicated on the spot. [42] More logging also renders forensic investigations after a security incident brutal since essential data may not be captured or recoverable [43]. That means organizations must implement more advanced monitoring solutions that suit the container ecosystem. This will be instrumental in guaranteeing better network visibility and control for offering threat detection and operational policy compliance.

### 3.2 Recommendations

- **Network Segmentation:**

Network segmentation is a critical approach in container environments, where the ultimate goal is to enhance security and efficiency. By segmenting the network into smaller, isolated portions, an organization decreases the attack surface, reducing the lateral movement of any potential threat within the environment [44]. Segmentation allows for more fine-grained access controls, ensuring that containers handling sensitive data operate in a secure environment while minimizing exposure to the less safe areas of the network [45]. Also, the implementation of network segmentation facilitates the process of compliance with regulatory requirements by simplifying the monitoring of data flows and ensuring that sensitive information is protected accordingly [46]. Finally, using network segmentation in containerized architectures enhances security and contributes to managing network resources more effectively.

- **Traffic Encryption:**

Such traffic encryption is a critical activity in ensuring that communications in the Docker container environments are kept safe from prying eyes through interception and unauthorized access. Containers are usually deployed in a dynamic and distributed system; hence, any application of encryption protocols like TLS ensures data in motion within the containers is private [47]. Traffic encryption would help protect the integrity of the data and reduce risks from man-in-the-middle attacks that could take advantage of vulnerabilities in containerized applications. In addition, using certificate management tools would further simplify the deployment of TLS certificates across Docker containers and hence improve the general state of security while ensuring adherence to regulations on data protection [47]. By prioritizing traffic encryption within their container orchestration strategies, organizations can enhance their defenses against potential security breaches while reinforcing the security of their applications.

- **Zero Trust Network Security Model:**

Zero Trust assumes excellent significance in containerized environments, as the dynamic nature of containers brings specific security-related challenges. The model operates on the principle of "never trust, always verify," where each request to access something is treated as if it could be malicious, no matter where on the network [48]. In container architectures, where microservices are in regular communication across untrusted networks, one must seriously identify and authorize the communications through proper identity verification and strict access controls to keep these interactions secure [49]. This allows organizations to monitor user behaviors and traffic patterns in real-time and to detect and respond to abnormalities that may indicate a security breach [50]. The reason is that integrating

Zero Trust into the current security frameworks will ensure sensitive data security both at rest and in transit, reducing risks that come through the lateral movement of threats in containerized applications [50]. Zero Trust opened the door to vast security posture improvement in container environments by allowing all components to undergo consistent assessment and protection against potential attacks.

- **Dynamic Policy Management:**

Dynamic policy management within container environments is an essential factor that guarantees security and operational policies will seamlessly adapt to rapid changes inherent in container orchestration. Due to the transient nature of containers, policies should be able to make real-time adaptations according to workload, resource availability, and security needs [51]. This calls for automating access control and resource allocation against real-time data, further enhancing security and efficiency [52]. In addition, dynamic policy enforcement allows an organization to apply compliance and governance more effectively to adapt to the varied operational contexts and dependencies forming across modern application architectures [53]. By taking advantage of machine learning and real-time monitoring utilities, containers can foster more intelligent decision-making to ensure that policies are reactive and proactive in dealing with future challenges and threats within the container ecosystem[54].

- **Network Monitoring and Logging:**

Network monitoring and logging in Docker container environments are crucial for ensuring both security and operational efficiency. As containers dynamically spin up and down, maintaining visibility across multiple instances becomes challenging, necessitating advanced monitoring solutions to track network activity effectively [55]. Implementing tools like the Round Robin Database (RRD) allows organizations to record key performance metrics, including memory usage, CPU percentage, and network throughput, facilitating timely data evaluation and visualization [56]. Furthermore, logging mechanisms integrated into orchestration platforms like Kubernetes enable detailed tracking of network interactions, essential for identifying anomalies and potential security threats. By establishing robust monitoring and logging practices, organizations can enhance their ability to respond to incidents in real time and uphold compliance with regulatory standards, ultimately strengthening their overall security posture within a containerized environment [57].

- **Firewalls and Intrusion Detection Systems (IDS):**

Firewalls and IDS in the Docker container environment will help protect against potential security threats. Firewalls help to securely segment containerized applications from unwanted incoming and outgoing traffic based on predefined security rules, thus offering high levels of protection against unauthorized access and attacks that are based on network-level exploitation [58]. Meanwhile, IDS solutions enhance security by constantly monitoring network traffic and container activities to detect and alert administrators about suspected behaviors or anomalies that could signify a security breach [58]. Moreover, implementing these security measures can facilitate compliance with regulatory standards, given that they provide logging and real-time alerts essential in auditing and forensic analysis [58]. By layering these with firewalls and IDS, the organization will strengthen prevention and detection while enabling responses to various security incidents in Docker environments, enhancing applications' overall security posture.

- **Microsegmentation:**

Microsegmentation is the security approach inside container environments, where it enhances application security by creating fine-grained network boundaries around workloads. It allows organizations to enforce specific security policies that govern container communications, reducing unauthorized access and constricting lateral movements across a network [59]. Through the isolation of different containers, microsegmentation does help narrow the attack surface, making it harder for attackers to leverage a vulnerability in one container to affect others [60]. This approach is instrumental in dynamic environments where containers are spun up and spun down regularly, as it enables runtime security policies with minimal performance impacts [61]. Ultimately, microsegmentation is critical in establishing a solid security posture in containerized applications and ensures that sensitive data is protected through its lifecycle.

## 4. Configuration Management: Importance of Configuration Management

Configuration management is the foundation for secure containerized deployments, ensuring systems are created, maintained, and updated organizationally. With the rise of containerized applications, secure configurations have become paramount in mitigating data leaks, privilege escalation, and unauthorized access. Configuration management allows an organization to maintain a consistent and safe environment throughout the entire SDLC, reducing the risk of human error while enforcing security best practices **(Mason & Kim, 2021).**

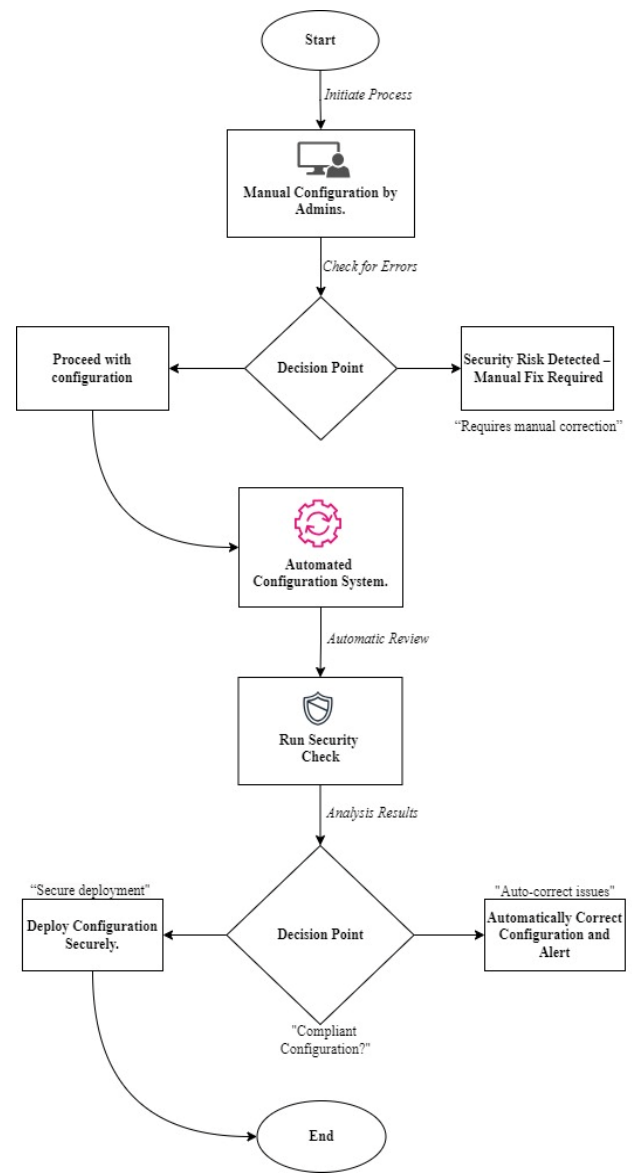

*Fig3. Automated configuration management and its impact on security*

This workflow illustrates the transition from a manual to an automated configuration management process, emphasizing the enhanced security and efficiency achieved through automation.

**1.Start:** The process begins with an initiation phase labeled as "Start."

**2.Manual Configuration by Admins:**

- o Initially, configurations are handled manually by administrators. This step is labeled "Manual Configuration by Admins," highlighting the reliance on human intervention. A user icon represents this stage to signify the involvement of personnel.

**3. Error or Misconfiguration? (Decision Point):**

- A decision point assesses if any errors or misconfigurations are present. If a misconfiguration is detected, the process diverges into a corrective action path. Otherwise, the configuration proceeds as intended.

- Yes Path: If an error is identified, it moves to "Security Risk Detected – Manual Fix Required," indicating that the issue must be manually resolved. The path to this step is marked with "Requires manual correction."

- No Path: If no errors are found, the workflow proceeds without further intervention under the label "Proceed with configuration."

**4. Automated Configuration System:**

- This step introduces the automated configuration management system, eliminating the need for continuous manual input. A gear or robot icon represents automation to denote the shift from manual handling to an automated system.

**5. Run Security Check:**

- The automated system conducts a security check on the configuration. This stage is represented by a shield or lock icon, symbolizing the security verification process inherent in the automation sequence.

**6. Compliant Configuration? (Decision Point):**

- Another decision point evaluates if the configuration is compliant with security policies.

- **Yes Path**: If compliant, the configuration is securely deployed, marked as *"Deploy Configuration Securely."* The flow to this deployment step is labeled "Secure deployment."

- **No Path:** If non-compliance is detected, the system moves to **"Automatically Correct Configuration and Alert."** Here, the automated system not only corrects the configuration but also sends an alert for review. The flow to this correction step is labeled *"Auto-correct issues."*

**7. End:**

- The process concludes with a secure and verified deployment, labeled as "End," ensuring a robust configuration without manual error risks.

*4.1 Current Gaps*

- **Default Configurations:**

While Docker default configurations help fast application setup and deployment, they can pose serious security risks if not correctly set. Although they enable a user to initiate the work, in most cases quickly, these default settings cut corners for ease at the cost of security, thus making possible misconfigurations vulnerable to be exploited by malicious actors [62]. Many users could only rely on default settings to understand their meanings and keep systems open to risks like unauthorized access and network vulnerabilities [62]. The most alarming discoveries of dependence on default configurations are those from industrial control systems and web servers since insecure settings can result in catastrophic failures or security breaches [63][64]. For instance, numerous users of Docker may refrain from modifying the default settings about container networking, which can inadvertently permit undesirable traffic and increase vulnerability to external threats.

Consequently, organizations ought to implement robust security practices, including the regular review and customization of these default settings, to enhance the security of their Docker environments and mitigate the risks associated with improper configurations [65]. Organizations can improve the overall security posture of their Docker deployments by prioritizing appropriate configurations rather than relying on default options.

- **Poor Secrets Management:**

Poor secret management inside Docker containers creates serious security risks in modern cloud-native applications, which rely on securely storing sensitive data, such as passwords, API keys, and encryption keys [66]. Traditional approaches include hardcoding secrets directly into images or environment variables, potentially leading to unintended exposure, thus complicating the protection of sensitive data [66]. Besides, with a central management solution, it's easier for an organization to control the distribution and access to secrets, thereby increasing unauthorized access and data breaches [66].

- **Lack of Security Baseline Enforcement:**

The lack of adequate security baseline enforcement within Docker containers can lead to some vast vulnerabilities that an organization would be forced to fix to maintain security. With security baselines set up, containers could become misconfigured, leak sensitive information, and increase the attack surface [67]. Without automation, security policies would not be consistently enforced across all containers, leading to varied levels of security that can be compromised by malicious actors [68]. More precisely, organizations risk deploying containers containing outdated or vulnerable components without correct baseline enforcement, further expanding the attack surface [69]. The situation is further aggravated by the dynamic nature of the containerized environments, where containers constantly get started and destroyed, hence requiring vigorous mechanisms to be laid down to ensure their real-time compliance with security best practices [70]. Lastly and most importantly, the inability of security baseline enforcement lies at the very core of all organizations using Docker containers; hence, this approach not only secures but also enables compliance and operational integrity.

- **Manual Configuration Changes:**

Manual Docker container configuration changes introduce security vulnerabilities and operational inefficiencies. Developers allowed to create or define configuration settings can create misconfigurations that would adversely affect the intended security posture for the containerized environment [71]. This is serious since containers run directly with the host kernel, opening up a larger attack surface if not managed carefully [72]. Besides, the lack of automation in the setup of Docker containers can result in environment inconsistencies, making it more difficult for teams to reproduce the setups across different stages of development and production [73]. For instance, errors made in changes to Dockerfile configurations or the interaction with Docker Compose can trickle into the deployment pipeline, causing operational hiccups and giving rise to technical debt buildup [74]. Such risks can be mitigated if the organization automates all configuration changes and standardization through Infrastructure as Code [75]. This will make the environments more secure with consistent configurations and more efficient regarding container deployment management [76]. Thus, automation reduces human errors that happen pretty frequently in organizations using a manual configuration process.

- **Inadequate Role-Based Access Controls (RBAC):**

Bad RBAC in Docker containers significantly increases the possibility of security compromise due to a lack of proper restrictions on sensitive resources. Most of Docker's default security mechanisms grant extensive privileges, which might allow users with host-level access to run and manipulate containers as root, possibly leading to unauthorized actions and data breaches if extended to unprivileged users [77]. This all-or-nothing access control model defeats the basic principle of least privilege, which states that users should have only the level of access necessary to accomplish their work[78]. More importantly, traditional RBAC systems being used in Docker environments may need to be more dynamic to adapt to the evolving nature of containerization, thus leaving potential gaps in access management due to changing users and roles[79]. Lacking robust RBAC, which also involves real-time auditing and modification of user permissions, entities risk internal attack and compliance violation[80]. In that respect, working out improved RBAC mechanisms within Docker to keep containerized applications safe from unauthorized access to specific resources remains highly important for maintaining the proper security posture across the infrastructure.

*4.2 Recommendations*

- **Secure Defaults:**

Setting secure defaults in Docker containers guarantees that the attack surface is minimized and provides security assurance for an application deployment. Configuring secure defaults minimizes the attack surface area, allowing an organization to reduce the risks of vulnerabilities concerning containerization [81]. This default configuration incorporates best practices intended to help protect against unauthorized access, data breaches, and other potential risks that might occur in containerized environments. Docker has complexities in terms of configuration issues by default due to the very design: Containers share the host operating system kernel. Such a design gives rise to privileged escalation and network attack risks if proper security is not in place. For example, the default networking could enable all containers to talk with each other without restriction, opening the possibility of sensitive data exposure if managed correctly. More robust network isolation with mandatory access control could better reinforce the defaults to protect container communications [82]. Besides, Docker image's system call and capability configurations should be precisely tuned according to the least privilege principle. Specialized tools like SysCap allow the automatic generation of secured configurations based on the calls and capabilities that specific images need[83]. With proper automation at the front, an organization can make security hardening easier while sustaining the best security posture against ever-evolving threats [83]. That said, using secure default Docker container configurations is critical for ensuring that applications are adequately protected. Other security best practices involve network isolation and customized system configurations to minimize the vulnerabilities of Docker's design. These measures ensure that a container environment is resilient and can withstand various security challenges.

- **Effective Secrets Management:**

Effective secrets management in Docker containers is about keeping sensitive information, such as passwords, API keys, encryption keys, and more, secure from potential unauthorized access and data breaches. A secure secrets management solution allows organizations to securely store and manage secrets outside the application code itself, thereby minimizing the risk of exposure resulting from misconfigurations or inadvertent leaks[84]. That would be the proper inclusion of mature tools like HashiCorp Vault or AWS Secrets Manager because both solutions provide rich functionality for secure storage, access, and rotation[85]. Another reason for using Docker's built-in secret management features is that it enhances security by injecting the secrets into containers at runtime rather than hardcoding them within images or setting them in environment variables[86]. This dynamic injection helps reduce the risk of secrets becoming exposed through logs or process listings, thereby fortifying security further. Additionally, an organization should implement strict access control to ensure that only specific services and human users can access particular secrets[87]. Regular auditing and real-time monitoring also play an essential role in identifying and mitigating potential risks associated with secret exposure[4]. Summing up, with the help of dedicated secrets management solutions, Docker's secret management capabilities, and strict access controls, an organization can reinforce its security posture and manage sensitive information effectively within Docker containers.

- **Establishing Security Baselines and Policies:**

Creating security baselines and policies in Docker containers is essential for the holistic security of container applications. Security baselines detail the minimum security requirements for Docker containers, thus forming a base to lay secure practices[88]. This will help organizations identify acceptable configurations and settings to reduce possible vulnerabilities that can be used by malicious actors[89]. To set up these baselines accordingly, organizations should thoroughly scan their container environments for potential risks associated with misconfigurations or insecure setups[90]. This includes cataloging container configurations, privileges, and network settings and determining what compliance standards are necessary and relevant to the organization [91]. Secondly, security policies must be implemented to control how Docker containers are created, deployed, and maintained throughout their life cycle. This includes guidelines on user access controls, image validation practices, and incident response protocols. Organizations can also leverage automated tools for continuous monitoring and compliance checks against the established security baselines, ensuring that deviations are promptly addressed[1]. In closing, security baselines and their implementation in Docker containers build a barrier that lessens the attack surface, allowing for compliance and proactive security mechanisms against threats that evolve every second in the container environment.

- **Automated Configuration Management and Orchestration:**

Automated Docker configuration management, therefore, is a crucial process in making deployments of containerized applications more effective, consistent, and secure. Using tools and protocols that enable automation in configurations allows an organization to drastically reduce risks from manual configuration, which often produces anomalies and vulnerabilities. Following Infrastructure as Code principles, automated configuration management enables teams to define, manage, and deploy configurations via version-controlled scripts, ensuring they are aligned with best practices for minimal error probability. Due to the integration of orchestration tools like Kubernetes, the advantages extend further, from the same benefits to facilitating automated deployment, scaling, and management of containerized applications across diverse environments in optimized operations that foster agility. Furthermore, this automation allows quick reaction to incidents because configurations can be changed or rolled back with negligible downtime, improving continuous integration and delivery practices crucial in today's rapid software development world[92]. Besides these operational efficiencies, automated configuration management in Docker strengthens security by persistently applying security patches, conducting compliance scanning, and reducing the attack surface created by insecure configurations[93]. More organizations move to microservices architectures and cloud-based deployments, so investments in automated configuration management tools are almost required to establish a robust, scalable infrastructure that will meet evolving business needs while ensuring high service availability and security levels.

- **Container Runtime Security:**

Container runtime security is all about protecting containerized applications at runtime, which not only includes the protection of the running container itself but also extends the host environment from various security attacks. In this security area, best practices will include strict access controls, constant monitoring of container activities, and automated security tools that can help prevent unauthorized access and privilege escalation [94]. The temporary nature of containers makes them even more vulnerable and, therefore, requires robust security that can adapt dynamically in real time[95]. Additionally, approaches like automated seccomp profiling can significantly enhance the security posture by defining and enforcing system calls that containers should be permitted to invoke, thus narrowing the attack surface[3]. Therefore, as organizations increasingly leverage container orchestration platforms like Kubernetes, adequate runtime security will become critical for maintaining deployed applications' overall integrity and safety [95].

- **Implement Role-Based Access Control (RBAC):**

Implementing Role-Based Access Control in Docker containers is essential for enhancing their security and ensuring users have appropriate access levels given their organizational roles and responsibilities. RBAC enables the administrator to define user roles and assign permissions based on the principle of least privilege; it minimizes the risk of unauthorized access to sensitive resources [97]. In a Docker ecosystem, this can be possible through orchestration platforms like Kubernetes since it has RBAC out of the box. Such orchestration platforms give fine-grained control over what a user can do against Docker resources like containers, images, and services[98]. RBAC enables organizations to enforce uniform security policies by building roles that encapsulate permission; hence, it is easy to manage users because the roles will be updated or reassigned to other users instead of updating resource permissions singularly [99]. RBAC implementation will also assist in attaining regulatory compliance since organizations will keep auditable access controls and enhance the general security posture in a containerized environment.

- **Immutable Infrastructure and Image Hardening:**

Immutable infrastructure and Docker image hardening are two more exciting topics that increase security and reliability in containerized applications. Conversely, immutable architecture refers to elements of the application architecture that cannot be changed after deployment but are replaced entirely by new versions when changes and updates are made[100]. This makes dependency management and configurations easier, reducing the chances of inconsistencies or vulnerabilities resulting from a system that changes over time. Complementary to this methodology, hardening Docker images involves best practices to secure images at build time by reducing the number of installed packages, scanning images for vulnerabilities, and running containers with non-root user permission[101]. Integrating immutable infrastructure with robust image-hardening practices gives organizations a much more secure and efficient deployment process. This approach ensures that applications operate in an identical, predictable state that minimizes attack exposure. The strength in such synergy pertains not just to security but also to the system's overall reliability, thus making the strategy mentioned above highly critical for modern cloud environments.

- **Implement Continuous Monitoring and Logging:**

Continuous monitoring and logging of Docker containers ensure security, performance, and reliability for containerized applications. With continuous monitoring, the tracking of containers about metrics, resource usage, and application performance happens in real time, thus allowing the organization to quickly identify and solve issues related to resource shortages or even a possible security breach [101]. Coupled with logging, this approach provides unparalleled details about container behavior for comprehensive audits and efficient troubleshooting should something go wrong[102]. Special monitoring tools and frameworks enable an organization to automate log collection and analysis, which helps meet regulatory requirements and expands visibility into the containerized environment[103]. Thus, Docker containers can ensure efficiency and safety by including continuous monitoring and extensive logging. This pays dividends in terms of swift incident response times and management. Docker containers' efficiency, security, and operational effectiveness determine a perfect selection of monitoring tools. Prometheus is one of the most popular open-source monitoring solutions, and it is very efficient in gathering and querying metrics from containerized applications and, therefore, most appropriate for dynamic environments[104]. Prometheus can be combined with Grafana so that users can build comprehensive dashboards to track real-time data and performance metrics[105]. On top of that, cAdvisor provides a good level of resource utilization and performance feedback by monitoring CPU, memory, and network usage for each container[106]. For security-focused monitoring, Wazuh and Sysdig provide extensive solutions to find anomalies and maintain compliance through log and container activity monitoring[107]. By baking these tools into a Docker management strategy, organizations will be better equipped to handle their containerized applications and make quicker detections of issues, thus raising overall system reliability.

- **Integrate Configuration Management with CI/CD Pipelines:**

It makes the integration of configuration management with Continuous Integration/Continuous Deployment pipelines for Docker containers a must-have in order to guarantee that applications deployed are consistent, reliable, and scalable. Thereby, it allows the immediate automated provisioning and configuration of containerized environments with minimal human intervention, which reduces configuration drift over time. The recommended toolsets to facilitate this integration include orchestration with Kubernetes, automation in configuration management with Ansible, and Jenkins, a viral CI/CD tool that works quite well with Docker[108]. In addition, GitLab CI and CircleCI provide excellent support for Docker container workflows, enabling fast deployments and easy rollbacks if something goes wrong in deployment[109]. These tools working in harmony help organizations to facilitate expedient deployment processes, ensure configuration compliance, and boost the overall efficiency of the software development life cycle.

## 5. Supply Chain Security: Importance of Supply Chain Security

Securing the software supply chain is key in maintaining the integrity within the development process. With the growing push toward reliance on third-party components and open-sourced software, the likelihood of introducing vulnerabilities through external dependencies has gone up several notches. A compromised supply chain could result in a data breach, unauthorized access, or the implementation of malicious software, thereby affecting the whole software life cycle in its course of development to deployment. (**Kim, Su Jin,2008**)

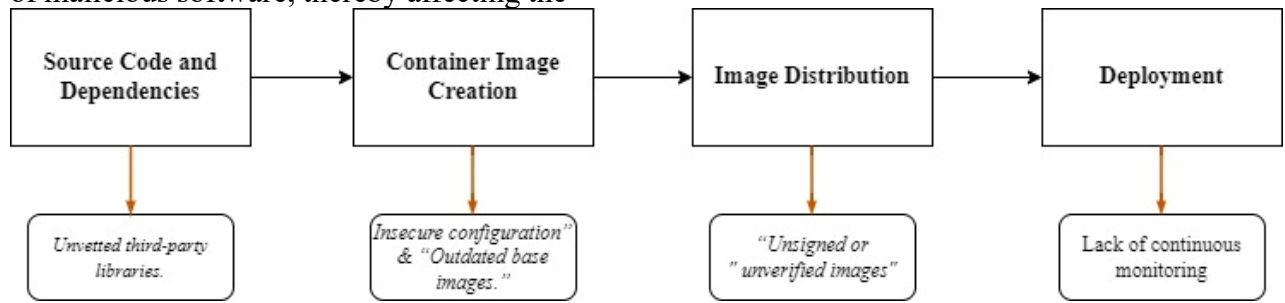

*Fig4 Supply Chain Management Gaps*

This diagram outlines the critical stages in the container supply chain, highlighting vulnerabilities at each stage that expose systems to security risks. Each stage and its associated security gaps are described below to clarify the potential threats in the container supply chain.

1. **Source Code and Dependencies**:

The supply chain begins with the source code and its dependencies. A major risk at this stage is the use of **unvetted third-party libraries**. Such dependencies, often sourced from public repositories, may contain malicious code or known vulnerabilities if not carefully reviewed. This step emphasizes the importance of thorough vetting to prevent introducing hidden security issues into the codebase.

2. **Container Image Creation**:

During the image creation process, several security gaps can arise. Common vulnerabilities include **insecure configurations** and **outdated base images**. Insecure configurations might expose unnecessary services or ports, increasing the attack surface. Similarly, outdated base images may lack crucial security patches, creating exploitable vulnerabilities. This stage underscores the need for regular updates and secure configuration practices.

3. **Image Distribution**:

At the distribution stage, images move from creation to deployment environments. A primary gap here is the reliance on **unsigned or unverified images**. Without proper image verification, it is challenging to ensure the authenticity and integrity of images, potentially allowing compromised or tampered images to enter production. Proper signing and verification practices are essential at this stage to safeguard the integrity of the container images.

4. **Deployment**:

The final stage, deployment, involves moving the image into production. **Lack of continuous monitoring** poses a significant risk at this phase. Without real-time monitoring, it is difficult to detect unexpected behaviors, unauthorized access attempts, or new vulnerabilities. Continuous monitoring and logging are essential to maintain a secure deployment environment and to respond quickly to any security incidents.

*5.1 Current Gaps*

- **Unverified Code Integrity:**

Unverified code integrity within Docker containers involves huge security risks, as the application of the host's operating system kernel by the containers results in poor isolation[1]. This weakness opens up avenues for attackers to influence the possible manipulation of these containers to consequence in issues such as container escape, where malicious activities can affect hosts and other containers and ultimately lead to instability in the security of the whole system[110]. To minimize these risks, it is important to apply some security hardening solutions that validate the integrity of container images before their deployment. This can be achieved by implementing image measurement methodologies that provide early detection of vulnerabilities, ensuring that images are unchanged and not tampered with throughout their life cycle[110]. Moreover, a container integrity measurement module can be used to verify key components, such as the code segment and shared libraries, hence enhancing the protection against unauthorized code execution[110]. Besides, strict access control and system whitelist development for container processes can substantially reduce the attack surface, thereby protecting both the containers and the host system from possible attacks[1]. Safety and reliability of Docker container usage in production environments could, therefore, be achieved only when unverified code integrity is addressed through rigorous security measures and continuous monitoring.

- **Outdated Dependencies:**

Outdated dependencies in Docker containers are a big pain for developers and organizations, as they have the potential to turn into major security vulnerabilities and application instability. In Docker's inheritance model, child images very often depend on parent images, which can contain a myriad of outdated components that might reach into the child images, further increasing the risk for exploits[111]. Studies have determined that, on average, about fifty percent of the child images using outdated parent images at the time of their creations had a typical lag of less than one month. By contrast, about seventy percent of the child images used outdated parent images when compared against the most recent version available, and the median lag was more than five months[111]. This means that users are required to carefully manage the provenance of their images and regularly update the dependencies of their containerized applications to keep their applications safe and running in a shifting security landscape.[111]

- **Inadequate Transparency in the Supply Chain:**

Lack of supply-chain transparency of Docker containers seriously jeopardizes the security and integrity of applications deployed in containerized environments. Since many layers of images and further dependencies often build containers, the lack of transparency regarding where such images originally came from and what changes have been made to them can lead to vulnerabilities and erode trust among stakeholders[112]. The obscured view may lead to challenges concerning compliance and accountability; this complicates organizations' abilities to trace any potential security issues back to their source[113]. Further, the various complicities in integrating third-party images raise the risks of introducing outdated or malicious code into applications, further exacerbating the problem of transparency[114]. Organizations should be able to reduce these risks by instituting broad monitoring and accountability frameworks and leveraging technologies like blockchain for increased visibility and traceability of the container supply chain[115].

- **Dependency Confusion and Typosquatting Attacks:**

The dependency confusion and typosquatting attacks in Docker containers are serious security risks, which may lead to harmful code injections and unauthorized access to sensitive applications. Dependency confusion refers to a situation where an attacker creates a malicious package in a private repository that has the same name as a valid package in a public repository, taking advantage of the system's attempt to resolve dependencies [116]. When a build process mistakenly pulls from the public repository, this malicious package can execute within the application and enable potential breaches and data theft [116]. Typosquatting, on the other hand, is where domain names or packages are registered that are slight misspellings of popular libraries or tools, and through which users will unwittingly download such malware instead of the real ones [116]. This attack factors in human mistake, where developers might omit noticing slight variations in the naming of packages while choosing their dependencies for the Docker containers [116]. The impacts from such kinds of attacks are very serious; they can eventually affect not just that single application but even the whole supply chain, creating a domino effect of vulnerabilities throughout systems that use those affected packages[116]. The only way that organizations can avoid such threats is by practicing good dependency management, where organizations utilize private registries, apply automated dependency scanning, and thoroughly vet third-party libraries prior to adding them into container images[116]. Indeed, good practice with regard to managing dependencies greatly reduces risks from Docker container dependency confusion and typosquatting attacks.

*5.2 Recommendations*

- **Code Signing:**

This is one of the major recommendations that go hand in hand with the security and integrity of Docker containers. Code signing leverages cryptography to sign Docker images and other artifacts created by the developer; thus, it delivers a mechanism that provides assurance about the authenticity of code integrity prior to deployment[117]. This practice ensures that production environments are used only with images that are trusted and not changed, hence reducing significantly the risk of a vulnerability via compromised or malicious code[118]. Besides, each organization should be in a position to have a full signing and verification process, hence using DCT for applying consistently the image signing policies across their respective pipelines for CI/CD [119]. Accordingly, by embedding code signing into the general methodology of container management, organizations improve their security posture in view of compliance with industry standards that engender trust among stakeholders and end-users.

- **Dependency Management:**

Effective dependency management in Docker containers is key in ensuring reliability, security, and performance for applications. This basically involves the identification, tracking, and resolution of interdependencies between the packages and libraries required by an application. For managing dependencies in Docker, a developer should start by using a well-structured Dockerfile, where they declare, with explicit naming, all the required dependencies during the build process. This is by specifying, in detail, base images that have all the software components necessary to avoid compatibility problems[120]. Besides, using tools like Docker Compose would add value in handling multi-container applications by defining and operating all services in one single YAML configuration file to make sure the dependencies are satisfied across a number of different environments[120]. This will go a long way in improving security by identifying and fixing vulnerabilities associated with out-of-date components[121]. Secondly, applying version control to Docker images and the `Dockerfile` will help teams track changes more effectively and support collaboration in teams. Thus, an organization is able to optimize dependency management inside Docker containers for continuous software delivery.

- **Software Bill of Materials (SBOM):**

A Software Bill of Materials is a standardized inventory of all components,

libraries, and dependencies that constitute a given software application[122]. In this case of Docker containers, whereby applications are commonly made from hundreds of third-party packages, an SBOM provides critical insight into the provenance of the software and integrity of its components[123]. This transparency helps an organization understand and manage vulnerabilities within its containerized applications because they can swiftly determine associated risks for each component[124]. Second, SBOMs assure regulatory requirements and the various industry standards for compliance by explicitly documenting the makeup of the software to make audits and security assessments easier[124]. With SBOMs in place, organizations are able to gain better security posture, ensure supply chain resiliency, and engender confidence with their stakeholders. The outcome is much safer and more reliable Docker deployments.

- **Implement Zero Trust in the Supply Chain:**

Moving toward the Zero Trust model within a Docker container supply chain is highly critical in terms of improving security and reducing vulnerabilities and threats-related risks. Such a model is based on the principle "never trust, always verify," implying that every request to access resources must be authenticated and authorized, whether from internal sources or from outside the organizational perimeter[125]. It means that organizations institute strict identity verification processes so that only trusted users and devices have access to important containerized applications and data, reducing the attack surface[125]. Additionally, continuous monitoring and logging of all activities with the deployment of containers provide real-time visibility for quick responses in cases of suspicious behavior or anomalies[125]. Proactive means a posture of improved security, regulatory, and industry standards that extend the Docker container ecosystem to be resilient against newly identified threats[125].

- **Combat Dependency Confusion:**

To effectively address dependency confusion in Docker containers, an organization should employ a set of measures aimed at enhancing security and ensuring the integrity of their applications. First, organizations must use private registries for their internal packages to avoid inadvertently downloading maliciously named public packages[126]. Implementation of naming conventions will make the internal packages well distinguishable from those outside, reducing possible conflicts[126]. Also, precise dependency constraint specification by a developer should be in place during the Docker build process to assure that only trusted package versions are used, avoiding any unintended updates that might introduce vulnerabilities[126]. Integrating automated security scanning tools into a continuous integration/continuous deployment pipeline allows for the early detection of vulnerabilities associated with dependencies and affords applications further protection from potential threats[126]. Ultimately, this would all be constantly watched and audited; hence, all unauthorized changes in the dependencies are instantly detected, fixed, and thus contribute to a safer container ecosystem[126].

## 6. Monitoring and Logging: Importance of Monitoring and Logging

Continuous monitoring and logging are essential components in preserving a secure containerized environment. In a context where containerized applications are subject to dynamic scaling, the possibility of security incidents occurring at any moment necessitates real-time detection as a priority. Monitoring identifies unusual behaviors and potential threats before escalation (IANS, 2022). Logging offers an audit trail, which is crucial for post-incident analysis, adherence to regulatory requirements, and proactive threat hunting. Collectively, they are the foundational element for a container security strategy that allows organizations to detect, investigate, and respond to incidents quickly. **(Mason & Kim, 2021; Pruitt, 2019).**

## 6.1 Current Gaps

- **Lack of Continuous Monitoring:**

Continuous monitoring is not possible inside Docker containers, which significantly affects application security, performance, and reliability. Without persistent visibility into container activities, organizations cannot identify real-time issues, which might extend to continued vulnerabilities, performance deterioration, and even complete outages[127]. By their very nature, Docker containers are ephemeral and dynamic; deploying monitoring solutions that can track changes and interactions within the container environment in near real-time is necessary. These measures increase the likelihood of security breaches going unnoticed, as unauthorized access and malicious activities may only be visible once the damage has already been done[128]. Besides, the lack of intensive monitoring complicates debugging activities, with increased consequences of more extended downtime and lowered operational efficiency. With continuous monitoring, an organization can maintain its DevOps efficiency, leading to slower deployment cycles and reduced confidence in the stability and security of containerized applications.

- **Fragmented Logging:**

The fragmentation of logs within Docker containers creates some application monitoring and debugging challenges. The transient nature of the containers explains this; logs often reside in different instances and may not persist when a container gets deleted. Such challenges can thus be related to losing crucial information that negatively impacts the capability to efficiently troubleshoot and perform any behavior analysis of an application over time[129][130]. Moreover, the diversity in log format from different containers complicates centralizing and analyzing log data. As more containers are added, the volume and variety of logs also increase, making meaningful insights challenging and time-consuming to extract. The inability to maintain a uniform logging approach across environments exacerbates the problem of fragmentation[129]. Logging fragmentation reduces operational efficiency and increases security risks since necessary logs that could explain security incidents may be ignored or inaccessible[130][129]. Organizations should employ central logging to mitigate this, where logs from all containers are presented through one interface. This would help improve analysis and monitoring to sustain the overall security posture of applications running within Docker environments.

- **Limited Context in Logs:**

Most Docker containers have very minimal context within the logs, making it hard for developers and operators to draw insightful conclusions from log data. This is because logs generated within a container might not include essential context about the container's environment, interactions between services, or the state of the infrastructure hosting it[131]. This could make the tracking of events in several different containers or services nearly impossible and the debugging difficult, which complicates an efficient incident response. Also, since logs are usually contained within containers, request flow tracking will be increasingly complex with a microservices architecture, which can lead to visibility gaps that cover up root causes. Overcoming these limitations involves adopting appropriate logging practices and tools, which can collate and enrich log data with relevant contextual information to enhance observability and operational effectiveness in containerized environments.

- **High Volume of Data:**

Managing vast volumes of data in Docker containers presents enormous challenges that could affect application performance and resource efficiency. As applications increasingly rely on containerized environments, the I/O demands associated with the data-intensive workloads quickly outstrip the underlying storage systems' capabilities, leading to bottlenecks. For example, although using high-performance storage solutions, such as NVMe SSDs, is bound to improve throughput, the concomitant execution of several Docker containers can still result in performance degradation. Indeed, application throughput can drop by as much as 50% compared with stand-alone applications if optimal configurations are not used[132]. This saturation makes effective strategies for managing volume imperative to allow applications to be responsive; poor resource allocation leads to extended latency and reduced efficiency. Second, with proper monitoring and tuning, organizations might be able to optimize resource utilization, which, in reality, defeats the very benefits that containerization is supposed to achieve in the first place. Overcoming these challenges will be crucial to maintaining high-performance levels while leveraging the benefits of containerized deployments in the modern cloud and data center infrastructure.

*6.2 Recommendations*

- **Comprehensive Monitoring:**

Complete monitoring in Docker containers is vital to guaranteeing high performance, security, and reliability for applications that work in a containerized environment. This type of monitoring provides real-time visibility into container health, resource utilization, and application performance metrics, thus offering an organization the required resources to identify and resolve potential issues even before they become serious[133]. The advantages of complete monitoring include almost unparalleled visibility into distributed systems, smoother troubleshooting, better allocation of resources, and more robust system resiliency in general[134]. The tools proposed to establish complete monitoring within the Docker environment are Prometheus for the functions of monitoring and alerting, Grafana for visualization of metrics data, and the ELK Stack: Elasticsearch, Logstash, and Kibana-in centralizing logs and performing analysis on them[135][136]. In addition, cAdvisor allows monitoring of resource usage and performance indicators of containers, while Kube-state-metrics provides critical insights into the health and status of Kubernetes-managed containers[137]. These tools will drive organizations to gain complete visibility over their containerized applications and empower them to achieve reliability, deploying more efficient incident response methodologies.

- **Centralized Logging:**

Centrally logging Docker containers is essential for managing and analyzing log data generated from containerized applications and services. It aggregates the logs from multiple sources such as containers, orchestration platforms like Kubernetes, and the host's operating system into a centralized repository for better accessibility and comprehensive analysis of log data[1]. Centralization of logs comes with such advantages as ease in troubleshooting, security monitoring, and regulatory compliance, given that the speed and ease of finding and fixing problems in distributed environments are higher[138][139]. Centralized logging also promotes better analysis for performance by relating logs to application behavior for good resource optimization and enhancement of overall system reliability.[140] Some of the recommended tools to implement centralized logging in Docker environments are the ELK Stack

Elasticsearch, Logstash, and Kibana, which efficiently aggregates and visualizes logs[3]. Meanwhile, the EFK Stack consists of Elasticsearch, Fluentd, and Kibana, which provide easy integrations of log data[140]. Graylog and Splunk are other good options, each providing advanced search features and real-time monitoring of logs that would assist an organization in finding issues before they occur and maintaining good operational health.

- **Automated Log Analysis:**

Log analysis with Docker containers is essential to enhance operation efficiency, thereby securing containerized environments. It allows for consistent collection, processing, and analysis of logs from the different services operating within a container to ensure an organization can uncover anomalies and fix issues faster, along with system performance monitoring[141]. The advantages of automated log analysis include reducing manual work, providing real application insights in real-time, and improving the correlation from distributed systems to root cause performance issues or security threats[141][142]. It's good regulatory practice, too, since the collection and analysis of log data are automated[141]. The ELK Stack, comprising Elasticsearch, Logstash, and Kibana, is highly recommended for advanced log aggregation and visualization[142]. Other vital tools are Fluentd for log collection and aggregation, Prometheus for monitoring, and Grafana for further data visualization[141][142]. Ultimately, these will continue to help an organization have deep visibility into its containerized applications for proper and timely decision-making.

- **Use of SIEM (Security Information and Event Management) Tools:**

SIEM systems must be introduced within Docker containers to monitor security incidents and manage incident responses in containerized applications. Employing SIEM solutions in Docker environments adds significant value in visibility into security violations, real-time analytics, and data correlation from a vast pool of sources such as container logs and orchestration platforms[143]. Generally speaking, the advantages of using SIEM in Docker containers are extended threat detection capability, compliance reporting efficiently, and better situational awareness about potential vulnerabilities[144][145]. Centralized logging and monitoring with the help of SIEM solutions enable organizations to rapidly identify and take action against security incidents to reduce overall deployment risk related to containers. Recommended tools for implementing SIEM in Docker environments include the Elastic Stack-ELK, which can aggregate logs and provide analytics[144]; Security Onion, which is a robust Linux distribution specifically designed for intrusion detection, network security monitoring, and log management[146]. Other significant tools include Splunk, a platform providing advanced data analytics and threat intelligence in a form that can easily integrate with container infrastructure[147]. Using these SIEM tools, it is possible to greatly enhance an organization's security posture in Docker container ecosystems.

- **Real-Time Alerting and Response:**

Real-time alerting and response in Docker containers are necessities for security and stability around containerized applications. Real-time monitoring enables an organization to detect abnormalities and potential threats faster so remediation actions can occur quickly before severe damage occurs[148]. This brings about increased advantages through heightened security postures, efficiency with incident response times, and reduced downtime- all to create genuinely resilient application environments[148]. Real-time alerting can automatically react to well-known threats, effectively reducing the need for human intervention while maintaining the consistency of security policy enforcement[1]. Wazuh is a highly recommended solution for real-time alerting and response in Docker containers. It's good at collecting log data, detecting malware, and automating active responses[148]. Wazuh can let it create an alert in case of a user or system change, providing complete monitoring, which is fundamental to the security and operation of cloud infrastructures and containerized applications.

- **Regular Log Audits and Compliance:**

Compliance with security regulations within Docker containers ensures data and application integrity in cloud environments. Compliance with recognized standards and regulations helps organizations minimize the risk of security breaches, data leaks, and service interruptions[149]. Compliance frameworks, including GDPR, HIPAA, and PCI-DSS, provide guidelines to help organizations establish appropriate security measures and controls for their Docker environments[150]. These are the benefits of maintaining regulatory compliance for security: increased trust by customers and stakeholders avoiding significant legal and financial penalties associated with non-compliance[151]. Moreover, best practices that various regulations require enable organizations to enhance their overall security posture and create a culture of accountability and due diligence about data protection[152]. In the end, security compliance plays a massive role in guiding the deployment and management of Docker containers, ensuring that security remains foremost in the whole application lifecycle.

## 7. Conclusion

In conclusion, this study highlights that while Docker containerization offers substantial benefits in scalability and operational efficiency, it also presents unique security challenges that must be proactively managed to ensure resilience in deployment environments. Previous research by[153] on container isolation and[154] exploration of container security tools underlines the need for robust isolation mechanisms and runtime protections to prevent unauthorized access and mitigate misconfigurations. As shown in these studies, a failure to address these aspects could expose containerized applications to significant vulnerabilities, which this research further details across areas like image security, runtime threats, network security, and configuration management.

Our findings stress the importance of adopting trusted sources for container images, conducting regular image scans, and avoiding dependencies on third-party libraries without proper vetting.[155] notably identified supply chain vulnerabilities in containers, with the research finding that approximately 70% of container issues stem from unverified third-party dependencies. In alignment with their insights, this study advocates for using multi-stage builds and rigorous scanning tools, such as Snyk[156] or Trivy[157], as effective measures to minimize security risks introduced by public repositories.

Moreover, runtime security emerged as a priority, with recommendations to enforce strict privilege settings and to adopt namespace and cgroup isolations. [158] emphasized runtime misconfigurations as a common vulnerability, and practical guidelines underscore the necessity of real-time monitoring to detect unauthorized behaviors. Implementing real-time monitoring and anomaly detection solutions can enhance visibility and allow for faster incident response, a crucial capability in dynamic containerized environments.

Network security remains another area of critical importance. Research by [159] on container orchestration underscored the vulnerabilities linked to network misconfigurations, especially in environments like Kubernetes. Following his recommendations, this study advocates for network segmentation and the application of secure communication protocols to protect sensitive services from lateral movement attacks. Logging and monitoring traffic flows within container networks also play a significant role in threat detection and operational integrity.

Inadequate configuration management can leave containers exposed to avoidable risks. Studies by [160] on secure configuration baselines and [161]on proactive security policy enforcement both advocate for establishing strong configuration practices to mitigate potential attack vectors. Reflecting these findings, this study calls for organizations to regularly review and update container configurations, implement automation for configuration management, and set up baseline policies that ensure consistent and secure deployments.

Our research supports a layered approach to container security, a practice also endorsed by [162], highlighting that a combination of tools and practices across the software development lifecycle is essential to build a robust security posture. This approach encourages organizations to prioritize a security-focused culture, enforce access controls, and maintain vigilance through continuous monitoring and updates to protect against emerging threats.

Ultimately, as container technology evolves, so too must our dedication to security. By applying the measures outlined in this study, organizations can strengthen their defenses against vulnerabilities, ensuring secure, efficient, and resilient containerized applications. Adapting to evolving security needs, as and both affirm, cannot be overstated, particularly as containers become an integral part of the modern application ecosystem. This commitment to innovation and vigilance in security practices will be pivotal in fostering a sustainable and secure foundation for future container deployments.